\documentclass[aps,pre,twocolumn,amsmath,amssymb,showpacs,floatfix]{revtex4}

\usepackage{graphicx}
\usepackage{subfigure}

\begin{document}

\title{Nonequilibrium interactions between ideal polymers and a repulsive surface}
\author{Raz \surname{Halifa Levi}}
\email{razhalifa@gmail.com}
\author{Yacov Kantor}
\affiliation{Raymond and Beverly Sackler School of Physics and
Astronomy, Tel Aviv University, Tel Aviv 69978, Israel}
\date{\today}

\begin{abstract}

We use Newtonian and overdamped Langevin dynamics to study long flexible
polymers dragged by an external force at a constant velocity $v$. The work
$W$ performed by that force depends on the initial state of the polymer and the details
of the process. The Jarzynski equality can be used to relate the non-equilibrium
work distribution $P(W)$ obtained from repeated experiments to the equilibrium
free energy difference $\Delta F$ between the initial and final states.
We use the power law dependence of the geometrical and dynamical characteristics
of the polymer on the number of monomers $N$  to suggest the existence of a
critical velocity $v_c(N)$, such that for $v<v_c$ the reconstruction of
$\Delta F$ is an easy task, while for $v$ significantly exceeding $v_c$ it
becomes practically impossible. We demonstrate the existence
of such $v_c$ analytically for an ideal polymer in free space and numerically
for a polymer which is being dragged away from a repulsive wall. Our results suggest
that the distribution of the dissipated work $W_{\rm d}=W-\Delta F$ in
properly scaled variables approaches a limiting shape for large~$N$.
\end{abstract}
\maketitle

\section{Introduction}
Equilibrium interactions between a single polymer and
a repulsive surface have been a subject of intensive study for
several decades \cite{Binder83,DeBell93,Eisenriegler1993} and
benefited from the relation between the statistical mechanics of
polymers and the general theory of phase transitions
\cite{deGennes79,Cardy84a,Grosberg94,Rubinstein03}.
Current experimental methods allow a detailed study
of biological macromolecules \cite{Zlatanova06,Leuba01}. In
particular, atomic force microscopy \cite{Binnig86,Morita02,Sarid94}
is an important tool in measuring force-displacement curves of
biomolecules, and reconstruction of their free energy and spatial
structure \cite{Bustamante2000}.

A long polymer held by its end at a distance $h$ from a repulsive flat
surface (wall), experiences an \textit{equilibrium} {\em repulsive} force
$f_{\rm{eq}}(h)$, i.e., to keep the polymer in place an external force
$f=-f_{\rm{eq}}$ towards the wall must be applied at the end of the polymer.
If $h$ is significantly larger than the microscopic
length scale $a$, such as monomer size or persistence length, but is
much smaller than the root-mean-square (rms) end-to-end distance $R$
of the polymer, then the expression for the force, at temperature $T$, has a
particularly simple form
\begin{equation}\label{eq:feq}
f_{{\rm eq}}(h)={\cal A}\frac{k_{\rm B}T}{h},
\end{equation}
where the dimensionless prefactor $\cal A$ can be related to the critical
exponents of the polymer \cite{Maghrebi11,Maghrebi12}. (For non-flat scale-free
repulsive surfaces, such as cones, the prefactor $\cal A$ depends on the surface
geometry \cite{Maghrebi11,Maghrebi12,AK_PRE91,HK_PRE92}). In many cases the
polymer size $R$ is related to the number of monomers $N$ by $R \approx aN^{\nu}$.
In particular, for an \textit{ideal} polymer, in which the interactions between
non-adjacent monomers are neglected, $\nu=\frac{1}{2}$, while for polymers in
good solvents $\nu \approx 0.59$ \cite{DeBell93}. Thus, the work $W$ performed
by an external agent while moving a polymer {\em slowly} away from a surface
at fixed $T$, as well as the free energy difference $\Delta F=F_{\rm f}-F_{\rm i}$,
between the final free energy of the polymer in free space $F_{\rm f}$ and the
initial free energy when the polymer is attached to the surface $F_{\rm i}$, is
\begin{equation}\label{eq:Weq}
W=\Delta F=\intop_{0}^{\infty}f(h)\mathrm{d}h\approx-\intop_{a}^{R}f_{{\rm eq}}(h)
\mathrm{d}h=-{\cal A}\nu k_{{\rm B}}T\ln N.
\end{equation}
For an ideal polymer near a flat surface ${\cal A} \nu = \frac{1}{2}$
\cite{Maghrebi11,Maghrebi12}. The negative sign reflects the need to
push the polymer {\em towards} the wall as we {\em slowly} move it away.

Equation \eqref{eq:feq} for the force and the resulting Eq.~\eqref{eq:Weq}
correspond to quasistatic motions. However, if the change is performed at
a finite rate, then the work $W$ of the external agent will depend on the
details of the experimental protocol, as well as on the microscopic initial
state of the system and the specific realization of thermal noise, if such
is present. Consider a situation where the initial state, such as a polymer
attached to the wall, corresponds to an equilibrium situation at temperature
$T$, i.e., is selected from a canonical ensemble. When an external agent follows
an experimental protocol and performs work $W$, the system reaches a new
non-equilibrium state, such as having a polymer far away from a wall.
If we proceed to equilibrate the system at temperature $T$, it settles into
a state  characterized by the free energy $F_{\rm f}$. Repeated non-equilibrium
experiments result in the work distribution $P(W)$. A remarkable
relation derived by Jarzynski \cite{Jarzynski97a,Jarzynski97b} relates the
{\em exact} distribution $P(W)$ to the change of the free energy $\Delta F$
between the final equilibrated state and the initial equilibrium state by
\begin{equation} \label{eq:JE}
\langle e^{-\beta W}\rangle= \intop_{-\infty}^{\infty}e^{-\beta W}P(W)
{\rm d}W = e^{-\beta\Delta F},
\end{equation}
where $\beta=1/k_BT$ and $\langle\cdot\rangle$ denotes averaging over the
initial states and over realizations of thermal noise, if such is present.

\begin{figure}[t!]
		\includegraphics[width=8truecm]{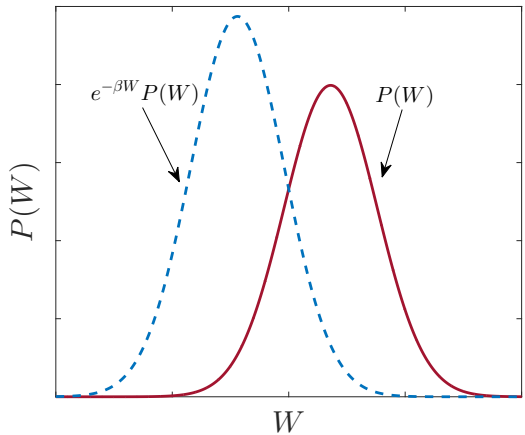}				
	\caption{(Color online) Illustration of the distribution of work
$P(W)$ (solid line) and the shifted function $G=e^{-\beta W} P(W)$
(dashed line). $P$ is measured by repeating the experiment, and $G$
needs to be reconstructed from the approximate knowledge of $P$.}
		\label{fig:shifted_P_W}
\end{figure}

At first sight, the Jarzynski equality (JE) provides a tool for easy calculation
of free energies from nonequilibrium measurements, and it has been used
to reconstruct free energies in certain situations
\cite{Liphardt02, Hummer05, Hummer05a, Harris07}. However, it has been observed
that for a system significantly out of equilibrium, the successful use of
Eq.~\eqref{eq:JE} requires an accurate knowledge of the probabilities of
nontypical (rare) events \cite{Jarzynski11}. (This can be explicitly observed
in the rare cases of exactly solvable systems, such as a one-dimensional Jepsen
gas  \cite{Lua04, Bena05}.) From the mathematical point of view, this happens
when the integrand of Eq.~\eqref{eq:JE}  $G(W)\equiv e^{-\beta W} P(W)$ has
a peak centered well below the position of the peak of $P(W)$, i.e., the
distance between the peaks exceeds the width of $P(W)$, as illustrated in
Fig.~\ref{fig:shifted_P_W}. In the latter situation, the function $G(W)$ is
reconstructed from the {\em tail} of $P(W)$, which cannot be accurately
estimated with a moderate number of repeated experiments. The separation
of $G$ and $P$ increases with departure from equilibrium in the experiment. A
convenient measure of this departure is the mean of the {\em dissipated work}
$W_{\rm d}\equiv W -\Delta F$. This $\langle W_{\rm d}\rangle$ vanishes
in quasistatic isothermal processes and increases with increasing rate of the processes,
and when it exceeds several $k_BT$ the free energy reconstruction becomes
unreliable. (It has been shown that the number of repeated experiments
required for a reliable reconstruction of $\Delta F$ increases exponentially
with $\langle W_{\rm d}\rangle$ of a {\em reverse} process \cite{Jarzynski06}.)
Thus, the borderline between easy measurements and practically impossible
ones is rather abrupt.

In this paper we consider the problem of a polymer, which is initially in
equilibrium near a flat repulsive wall, and is being dragged away with a
constant velocity $v$. The final state is when the polymer is in equilibrium
far away from the wall, such that we can treat it as being in free space.
The dynamics of the system will be either overdamped Langevin dynamics, in
which the inertia term is neglected, or Newtonian dynamics in which friction
and thermal noise are absent. We argue that there is a critical pulling
velocity $v_c$, such that for $v<v_c$ reconstruction of $\Delta F$ is
possible by using JE, while for $v>v_c$ such reconstruction is practically
impossible. In our discussion we will focus only on these two extreme cases,
and we will not consider the range of velocities around $v_c$ for which the
ability of reconstruction strongly depends on the number of experiments.
In Sec.~\ref{sec:Determining v_c in polymers} we present a heuristic argument
for calculating $v_c$ in rather general circumstances. In the remainder of
the paper, we provide supporting evidence for the analytically solvable case
of ideal polymers in free space (Sec.~\ref{sec:Gaussian chain in free space})
and for the numerically solved case of an ideal polymer near a flat repulsive
surface (Sec.~\ref{sec: polymer near a wall}). In Sec.~\ref{sec: Summary} we
summarize our results, and discuss their possible generalizations. Since our
work relies on the known results of a dragged harmonic oscillator, we provide
a short summary of these properties in Appendix~\ref{APPEN:A}.

\section{Critical dragging velocity: A heuristic argument}
\label{sec:Determining v_c in polymers}
Consider a situation in which a large polymer is being dragged away from a
repulsive wall by moving its end monomer at a constant velocity $v$. In the
initial (equilibrium) state the end monomer is attached to the wall, and in
the final state the end-monomer is at a distance significantly exceeding the
polymer size $R$, so that interactions with the wall can be ignored.

Many equilibrium properties of polymers have a simple power-law dependence on
the number of monomers $N$. In many cases dynamic features also have that property \cite{deGennes79,Grosberg94,Rubinstein03}. We would like to take advantage
of these scaling properties of polymers to estimate the critical velocity $v_c$,
which defines the borderline between ``fast" and ``slow" dragging.

Let us first consider the simple case of a polymer prepared in thermal
equilibrium, and subsequently disconnected from the thermal bath, i.e., its
subsequent motion will be determined by {\em Newtonian dynamics} (ND). For a polymer
at equilibrium, the velocity of its center of mass is
$\mathbf{v}_{\mathrm{cm}}=\frac{1}{N}\sum_{i=1}^{N}\mathbf{v}_{i}$, where
$\mathbf{v}_i$ is the velocity of the $i$th monomer. For a polymer at
equilibrium the average  $\left\langle \mathbf{v}_{\mathrm{cm }}\right\rangle =0$.
However, the typical or the rms velocity is
\begin{equation}
v_{\mathrm{cm}}=\sqrt{\frac{1}{N^{2}}\sum_{i,j=1}^{N}\left\langle \mathbf{v}_{i}\cdot\mathbf{v}_{j}\right\rangle }=\frac{v_\mathrm{th}}{\sqrt{N}},
\end{equation}
where $v_{\mathrm{th}}=\sqrt{d/ \beta m}$ is the rms thermal velocity of a
single monomer, while $m$ is the mass of the monomer, and $d$ is the spatial
dimension. [In further (approximate) calculations we will omit $d$.] The time $t$
it would take the polymer to move a distance equal to its own size $R=aN^{\nu}$ would be
\begin{equation} \label{eq:t_0 ND}
t=R/v_{\rm{cm}}\approx a\sqrt{\beta m}N^{\nu+1/2}.
\end{equation}
We expect that this will also be the time scale of the slowest internal motion
of the polymer. It is natural to define a ``slow motion" velocity $v$, such that
during the time $t$ the polymer is not dragged more than $R$, i.e., we must
require $v<v_{\mathrm{cm}}$. In other words, for the ND the borderline critical
velocity $v_c$ coincides with the typical velocity of the center of mass
$v_{\mathrm{cm}}$, or
\begin{equation} \label{eq:v_C Newton_Qual}
v_c\approx \dfrac{1}{\sqrt{\beta m}}N^{-1/2}.
\end{equation}
The ND approach neglects the interactions of a polymer with the surrounding
fluid and therefore its practical usefulness is limited. However, it presents a
theoretically important case that formed an essential part of the original
proof of JE \cite{Jarzynski97a,Jarzynski97b}, and provides important
insights into the relations between  the ``regular" mechanics and the thermal
physics. It also can be viewed as a limiting case of a general Langevin equation.

Alternatively, we can consider motion of the polymer in a very viscous fluid,
where the inertia can be neglected on sufficiently long time scales. In this
example we neglect hydrodynamic interactions, i.e., we consider the
``free-draining" \cite{DoiM.Edwards1986} regime. Such motion can be described by the
{\em overdamped Langevin dynamics} (OLD). In the overdamped regime, the center of mass
of an $N$-monomer polymer in free space, performs diffusion characterized by
a diffusion constant $D$, which is $N$ times smaller than the diffusion constant
$D_0$ of a single monomer. Therefore, the time $t$ it takes it to diffuse a distance
$R=aN^{\nu}$, is
\begin{equation} \label{eq:t_0 OLD}
t \approx \dfrac{R^2}{D} =\dfrac{a^2 N^{2\nu}}{D_0/N}= \frac{a^2}{D_0} N^{2\nu+1}.
\end{equation}
This is also the slowest relaxation time of an internal mode of the polymer \cite{deGennes79,DoiM.Edwards1986}.

If the polymer is being dragged with a velocity $v$, we would consider such
motion ``slow" if during the same time $t$, the distance $vt$ that the
polymer is dragged does not exceed $R$. This means that we need to have
$v<\frac{D_{0}}{a}N^{-1-\nu}$, or
\begin{equation} \label{eq:v_c wall langevin}
v_c \approx \frac{D_{0}}{a}N^{-1-\nu}.
\end{equation}

When hydrodynamic interactions are important (the Zimm regime
\cite{Zimm56,deGennes79}), a long polymer is not ``transparent" to the
surrounding liquid, and can be treated as a sphere
of radius $R$ diffusing in a liquid of viscosity $\eta$ \cite{deGennes79}
with a diffusion constant $D\approx 1/\beta\eta R$, where we omitted
a dimensionless prefactor. The time it takes for such a polymer to
diffuse a distance $R$ is $t\approx R^2/D\approx \beta\eta R^3$, which
leads to
\begin{equation} \label{eq:v_c hydro}
v_c \approx\frac{k_BT}{\eta R^2}\approx \frac{k_BT}{\eta a^2}N^{-2\nu}.
\end{equation}

The arguments presented in this section are equally valid for a polymer in
free space or near a repulsive wall, because in both cases the polymer will
have similar relaxation times.

\section{Gaussian polymer in free space}\label{sec:Gaussian chain in free space}

The arguments presented in the previous section were valid for a broad class of
polymer types and interactions between monomers. In this section we consider
a simple model of an ideal polymer, in which we neglect the interactions between
non-adjacent monomers of the chain, that are usually important in good solvents.
The model only retains the linear connectivity of the monomers and is analytically solvable.
Ideal polymers rarely represent experimental systems, but the scaling properties
of their static and dynamic characteristics provide guidance to the treatment of
more realistic and complicated models~\cite{deGennes79}.

\begin{figure}[t]
\includegraphics[width=8truecm]{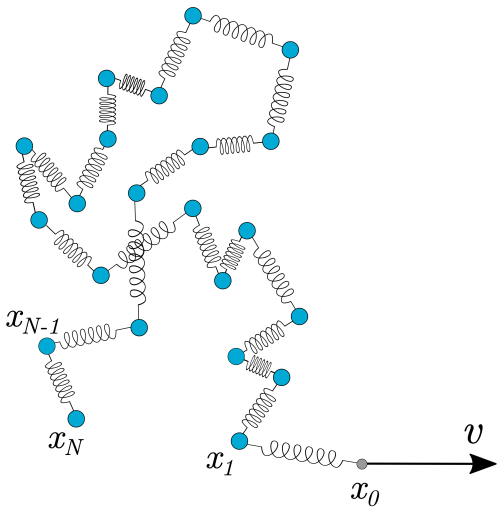}
\caption{(Color online) A beads-and-springs model of a Gaussian (Rouse) chain
is being dragged in free space by
pulling its end point $x_0$ with a constant velocity $v$, such that $x_0=vt$.
Note, that we consider only a one-dimensional dynamical problem, and the
second dimension in this figure is for illustration purpose only.}
 \label{fig:Polymer in free space}
\end{figure}

Consider a linear chain of $N$ identical monomers of mass $m$ connected
by  springs with constants $k$, such that the potential energy is given
by $\frac{1}{2}k\sum_{i=1}^{N}\left(x_{i}-x_{i-1}\right)^{2}$, where $x_i$
($i=1,...,N$) are the positions of the monomers, while $x_0$ is the position
of the end point to which the  first monomer is connected, as depicted in
Fig.~\ref{fig:Polymer in free space}. Such an energy describes the Gaussian
model of an ideal polymer, which in the polymer literature is well known
as Rouse model \cite{DoiM.Edwards1986}, although the latter term is also
used to describe the type of dynamics, rather than the polymer structure.
(The term ``Gaussian" refers to the
functional shape of the Boltzmann weight of this energy.) The microscopic
length $a$ is given by the rms separation between two consecutive monomers,
i.e., $a^2\equiv\left\langle \left(x_{i}-x_{i-1}\right)^{2}\right\rangle =1/\beta k$.
We can consider motion in three-dimensional space, with monomers positioned
at ${\bf r}_{i}$. However, the particular form of the potential
$\frac{1}{2}k\left({\bf r}_{i}-{\bf r}_{i-1}\right)^{2}$ splits into three
independent parts and the motions in different space directions are independent.
Thus, the only non-trivial part of the problem is in the direction
parallel to the velocity with which the point at ${\bf r}_{0}$ is being
dragged. This reduces the problem to a single space dimension.

The problem of {\em stretching} a Gaussian chain, when $\Delta F$
increases as a result of the external work, has already been
solved for OLD, and the distribution of work has been calculated
analytically \cite{Dhar05, Speck05}. We apply the same technique to
a problem with slightly different boundary conditions, for both ND
and OLD cases, and have a different goal: We consider the particular
case of dragging the polymer in free space, where the free energy
difference $\Delta F$ between the equilibrated final and initial
states vanishes. In this section we find analytical expressions for
$P(W)$, as well as $v_c$. The ND and OLD cases can be viewed as two
extremes of the general Langevin equation for the system.

\subsection{Analytical calculation of $P(W)$}

In the absence of friction and thermal noise, the equation of motion of
the $n$th monomer ($1\leq n \leq N-1$) is governed by the ND equations
\begin{equation} \label{eq:EOM_ND_Polymer}
\ddot{x}_n=-\omega^2 \left(2x_{n}-x_{n+1}-x_{n-1}\right),
\end{equation}
and for $n=N$\begin{equation}
\ddot{x}_{N}=-\omega^2(x_{N}-x_{N-1}),
\end{equation}
where $\omega\equiv\sqrt{k/m}$. It is more convenient to work in a reference
frame which moves with $x_0$, i.e., with a constant velocity $v$ such that
the position of the $n$th monomer (in the moving system) is
$\tilde{x}_n=x_n-vt$. In this reference frame the equations of motion can
be separated into $N$ independent (Rouse \cite{DoiM.Edwards1986}) eigenmodes
by decomposing the position of the $n$th monomer into its discrete Fourier
components,
\begin{equation} \label{eq:DST x_n}
\tilde{x}_{n}  =  A\sum_{q=1}^{N} \tilde{x}_{q}\sin\left(\alpha_{q}n\right),
\end{equation}
where $\tilde{x}_q$ is the amplitude of the $q$th mode, and
$A=\sqrt{\frac{2}{N+\frac{1}{2}}}$, while
$\alpha_{q}=\frac{\pi(q-\frac{1}{2})}{N+\frac{1}{2}}$ was chosen to satisfy
the boundary conditions, where one end of the polymer is fixed ($\tilde{x}_0=0$)
and the other end ($\tilde{x}_N$) is free. The equations of motion remain the
same in the moving system, with $x$ replaced by $\tilde{x}$. Substituting
Eq.~\eqref{eq:DST x_n} into Eq.~\eqref{eq:EOM_ND_Polymer} produces the
equation of motion for the $q$th eigenmode in the moving reference frame,
\begin{equation} \label{eq:EOM_ND_eigenmode}
\ddot{\tilde{x}}_{q}=-4\omega^2\sin^2\left(\alpha_q/2\right)\tilde{x}_q,
\end{equation}
which is a simple harmonic oscillator with frequency $\omega_q$ defined by
\begin{equation}\label{eq:w_q}
\omega_q^2\equiv 4\omega^2\sin^2\left(\alpha_q /2\right).
\end{equation}

The {\em constant} pulling velocity can be represented (for any
$n=1,\dots,N$) as
\begin{equation}
v=A\sum_{q=1}^{N}v_{q}\sin\left(\alpha_{q}n\right).
\end{equation}
In this particular case of constant pulling velocity, $v_q$ can be
simply expressed via the inverse transform as
$v_q=A\sum_{n=1}^Nv\sin\left(\alpha_qn\right)=\frac{1}{2}Av\cot(\alpha_q/2)$
and used to transform the solution back to the laboratory frame,
\begin{eqnarray}
x_{n} & = & \tilde{x}_{n}+vt\\
 & = & A\sum_{q=1}^{N}\tilde{x}_{q}\sin\left(\alpha_{q}n\right)
 +A\sum_{q=1}^{N}v_{q}\sin\left(\alpha_{q}n\right)t\\
 & = & A\sum_{q=1}^{N}\underset{x_{q}}{\underbrace{\left(\tilde{x}_{q}
 +v_{q}t\right)}}\sin\left(\alpha_{q}n\right).
\end{eqnarray}
Here, we defined $x_q=\tilde{x}_q+v_qt$, and the equation for the
$q$th eigenmode in the laboratory frame is given by
\begin{equation}
\ddot{x}_q=-\omega_{q}^{2}\left(x_q-v_qt\right).
\end{equation}
This can be viewed as $N$ independent simple harmonic oscillators with
frequencies $\omega_q$, being pulled by effective velocities $v_q$. Note
that, for large $N$, the frequency of the lowest mode
$\omega_{q=1}\sim\omega N^{-1}$ corresponds to the time it would take the
polymer to move a distance $R$, as in Eq.~\eqref{eq:t_0 ND} with $\nu=1/2$.

We now examine the other extreme of this problem, in which the polymer is
moving in a very viscous fluid, so that the inertia term can be neglected,
and its motion is described by OLD. The equation of motion of the $n$th
monomer, for $1\leq n \leq N-1$, is given by
\begin{equation} \label{eq:EOM_OLD_Polymer}
\gamma\dot{x}_{n}=-k\left(2x_{n}-x_{n+1}-x_{n-1}\right)+\eta_{n}(t),
\end{equation}
and for $n=N$
\begin{equation}
\gamma \dot{x}_{N}=-k\left(x_{N}-x_{N-1}\right)+\eta_{N}(t),
\end{equation}
where $\gamma$ is the friction coefficient and $\eta_n(t)$ is the thermal
noise associated with the $n$th monomer. The thermal noise is chosen to be
white Gaussian noise which satisfies $\left\langle \eta_n(t)\right\rangle =0$ and
$\langle \eta_{n}(t)\eta_{n'}(t')\rangle
=2\gamma k_{{\rm B}}T\delta(t-t')\delta_{nn'}$. The same decomposition that
was applied to the position $x_n$ and the pulling velocity $v$ in the ND
case, can be applied in this case too, while the decomposition of the noise is
\begin{equation}
\eta_{n}(t)=A\sum_{q=1}^{N}\eta_{q}(t)\sin\left(\alpha_{q}n\right),
\end{equation}
where $\eta_q(t)$ is the effective thermal noise acting on the $q$th
eigenmode, which satisfies
$\langle \eta_{q}(t)\rangle =0$ and
$\langle \eta_{q}(t)\eta_{q'}(t')\rangle =2\gamma k_{B}T\delta(t-t')\delta_{qq'}$.

Similarly to the ND case, the system is decomposed into $N$ independent
(Rouse) eigenmodes, where each one represents an independent overdamped
harmonic oscillator that is being dragged with an effective velocity $v_q$.
Each eigenmode satisfies
 \begin{equation} \label{eq:u_q_EOM}
\dot{x}_{q}=-\frac{1}{\tau_{q}}\left(x_{q}-v_{q}t\right)+\frac{1}{\gamma} \eta_{q},
\end{equation}
where \begin{equation}\label{eq:tau_q}
\tau_q\equiv\frac{\tau}{4\sin^2\alpha_q/2}
\end{equation}
is the relaxation time of the $q$th eigenmode, and $\tau \equiv \gamma /k $.
As we can see, the largest relaxation time $\tau_{q=1} \sim \tau N^2$ (for
large $N$) coincides with the time it takes the center of mass of the polymer
to diffuse a distance $R$ [Eq.~\eqref{eq:t_0 OLD}]. During the time $\tau$ a
monomer moves an approximate distance $a\approx\sqrt{D_0 \tau}$, where
$D_0=k_{\rm{B}}T/\gamma$.

Both in the ND and OLD cases we can treat the system as $N$ independent
harmonic oscillators, and write the total work $W$ done on the system
(by an external agent) during the pulling, as a sum of works $W_q$ done
on each single effective oscillator, i.e.,
\begin{equation}
W=\sum_{q=1}^{N}W_{q}.
\end{equation}

\begin{figure}[t]
\includegraphics[width=9truecm]{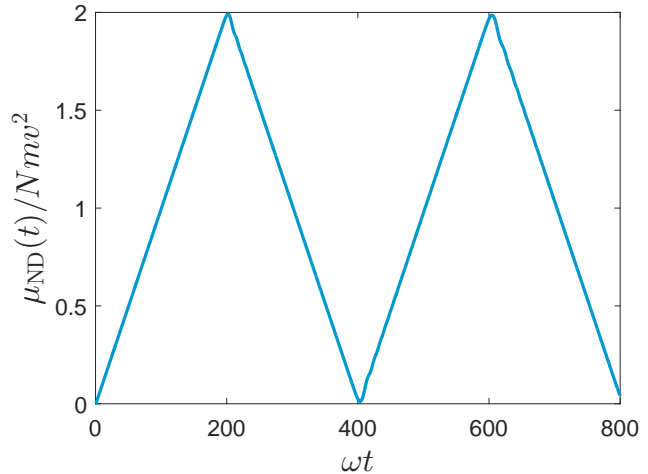}
\caption{(Color online) Normalized mean work $\mu_{\rm ND}$  for $N=100$
as a function of time $t$ (see text). For large $N$, $\mu_{\rm ND}$ resembles
a triangular wave with a period $T=4N/\omega$. (For large $t$ the apparent
periodicity disappears.)}
 \label{fig:mu_ND}
\end{figure}

Each $W_q$ has a Gaussian distribution with mean $\mu_q$ and variance
$\sigma_q^2$, such that
$\mu_q=\frac{\beta}{2}\sigma_q^2 = 2mv_q^{2}\sin^{2}\left(\omega_q t/2\right) $
for the ND case, and
$\mu_q=\frac{\beta}{2}\sigma_q^2=\gamma\tau_q v_q^{2}
\left(e^{-t / \tau_q}+t / \tau_q-1\right)$ for the OLD case (as shown in the
Appendix). Therefore, $W$ also has a Gaussian distribution
characterized by mean $\mu=\sum_{q=1}^{N}\mu_{q}$ and variance $\sigma^{2}=\sum_{q=1}^{N}\sigma_{q}^{2}$.

In the ND case the mean work is
\begin{equation} \label{eq:mu_ND}
\mu_{\rm{ND}}(t)=\frac{\beta}{2}\sigma^{2}_{\rm{ND}}(t)= \sum_{q=1}^{N} 2m v_{q}^{2}\sin^{2}\left(\frac{\omega_{q}t}{2}\right).
\end{equation}
For small $q$ the frequencies $\omega_q$ have almost integer ratios with
$\omega_{q=1}$ and, not surprisingly, $\mu_{\rm{ND}}(t)$ is ``almost" a
periodic function. For $N\gg 1$, it looks as a triangular wave depicted
in Fig.~\ref{fig:mu_ND} of amplitude $2Nmv^2$ with period
$T=2 \pi/\omega_{q=1}\approx 4N/\omega$. However, higher frequencies have
more complicated dependence on $q$ [see Eq.~\eqref{eq:w_q}], and after many
oscillations the appearance of the periodicity vanishes.

\begin{figure}[t]
\includegraphics[width=9truecm]{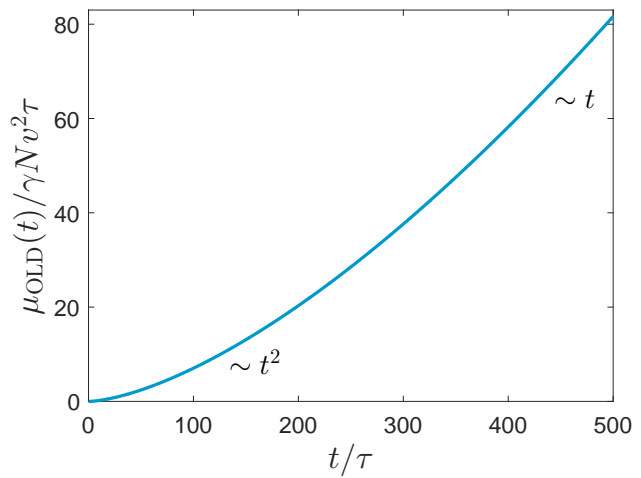}
\caption{(Color online) Normalized mean work $\mu_{{\rm OLD}}$ as  function of time $t$
for $N=100$. (See text.) At short times,
$\mu_{{\rm OLD}}$ is parabolic, and for large times it grows linearly with $t$.}
 \label{fig:mu_OLD}
\end{figure}

It is interesting to note that in terms of dimensionless variable $y\equiv \beta W$,
the probability distribution of the work $W$ has a very simple form,
\begin{equation}\label{eq:Preduce}
\tilde{P}(y)=\frac{1}{\sqrt{4\pi\tilde{\mu}}}
\exp\left[-\frac{(y-\tilde{\mu})^2}{4\tilde{\mu}} \right],
\end{equation}
where we used the relation \eqref{eq:mu_ND} between the mean and the variance
and the reduced mean $\tilde{\mu}\equiv\beta\mu_{\rm ND}$. For large $N$ and moderate
times it is convenient to express $\tilde{\mu}$ via triangular wave function $S$
(as in Fig.~\ref{fig:mu_ND}) of
unit amplitude and period leading to $\tilde{\mu}=2\beta Nmv^2S(t/T)$.
If we measure the pulling velocity in units of $v_c$ [as defined in
Eq.~\eqref{eq:v_C Newton_Qual}], i.e., $u\equiv v/v_c$,
and the total pulling length $L$ in units of polymer size $R$, $\ell\equiv L/R$,
then the expression for the reduced mean further simplifies to
\begin{equation}\label{eq:Mu_reducedND}
\tilde{\mu}=2u^2S(\ell/4u),
\end{equation}
where we used the fact that $R=aN^{1/2}$ and the mean separation between the
monomers is $a=1/\omega\sqrt{\beta m}$.

In the OLD case the mean work is
\begin{equation} \label{eq:mu_polymer_langevin}
\mu_{\rm OLD}(t)=\frac{\beta}{2}\sigma^2_{\rm OLD}(t)=\sum_{q=1}^N\gamma\tau_qv_q^2
\left(e^{-\frac{t}{\tau_q}}+\frac{t}{\tau_q}-1\right),
\end{equation}
which is depicted in Fig.~\ref{fig:mu_OLD}. For short times ($t\ll\tau_{q=N}\approx\tau$),
it increases parabolically with time: $\mu_{{\rm OLD}}(t) \approx \frac{1}{2}\frac{\gamma}{\tau}v^{2}t^{2}=\frac{1}{2}k\left(vt\right)^{2}$.
This corresponds to an external force stretching a single spring of the first
monomer connected to $x_0$, since during time $t<\tau$ only $x_0$ moves,
and the rest of the system ``does not know'' yet that it is being pulled.
In the long time regime ($t\gg\tau_{q=1} \approx \tau N^2$) the mean work
grows linearly with time as: $\mu_{{\rm OLD}}(t) \approx \gamma Nv^{2}t$.
This represents the work against friction performed by dragging $N$ monomers
together. (Other eigenmodes are already relaxed in the system.)

\subsection{Analysis of the results}

The JE in Eq.~\eqref{eq:JE} can be cast in the form of a cumulant expansion
\cite{Jarzynski97a,Hummer2001}:
\begin{equation} \label{Eq:cumulant exp}
\Delta F=-\frac{1}{\beta}\ln\left\langle e^{-\beta W}\right\rangle = \mu-\frac{\beta}{2}\sigma^{2}+...
\end{equation}
If $P(W)$ is a Gaussian, as in the case of our model in free space, this
expansion terminates at the second term. In addition, in free space
displacement of the polymer does not modify its free energy, i.e.,
$\Delta F = 0$. From Eq.~\eqref{Eq:cumulant exp} we conclude that in free
space $\mu=\frac{\beta}{2}\sigma^2$, which coincides with the analytical
results obtained by a direct calculation in the previous subsection.
In the particular case of Gaussian $P(W)$, $G(W)\equiv e^{-\beta W} P(W)$
is another Gaussian shifted by $2\mu$ towards lower values of $W$.
For slow motion we must have $\mu < \sigma$, i.e., the mean work (and the
shift between $P$ and $G$) does not exceed the width of the distribution.
At the critical velocity this relation becomes an equality. Due to the
relation between $\mu$ and $\sigma$, this condition becomes
\begin{equation} \label{eq:condition on mu}
\mu\approx 2k_{\rm{B}}T.
\end{equation}

In the case of ND, the mean value of work is bounded by $2Nmv^2$, and
therefore, from Eq.~\eqref{eq:condition on mu} we find that
$v_c\approx N^{-1/2}/\sqrt{\beta m}$, which is exactly $v_c$ that we found
in Eq.~\eqref{eq:v_C Newton_Qual}.

In the OLD case, $\mu_{\rm{OLD}}$ increases monotonically with time, making
the free energy reconstruction more difficult as $t$ grows. We would
like to drag the polymer a distance at least equal to its size, i.e.,
$L=v t \sim a\sqrt{N}$. (In free space it is a somewhat arbitrary choice, but
in the next section we will consider a polymer being dragged away from a wall,
and then such a choice of distance becomes crucial.) For such $L$ the condition
in  Eq.~\eqref{eq:condition on mu} translates into
$a \gamma N^{3/2}v_c\approx k_{\rm{B}}T$,
which defines the critical velocity,
\begin{equation} \label{eq:critical velocity OLD}
v_c\approx \frac{1}{a \gamma \beta}N^{-3/2}.
\end{equation}
Substituting $a\approx\sqrt{D_0 \tau}$ brings us back to the same critical
velocity  that was found in Eq.~\eqref{eq:v_c wall langevin}, with $\nu=1/2$.
In terms of dimensionless variable $y=\beta W$ the probability distribution
of work is given by Eq.~\eqref{eq:Preduce}, with reduced mean
$\tilde{\mu}\equiv\beta\mu_{\rm OLD}$. For times larger than the relaxation time
of the polymer and for $N\gg1$, we get $\tilde{\mu}=\beta\gamma Nv^2t$, which can
be conveniently expressed via relative distance $\ell=L/R$ and relative
velocity $u=v/v_c$, where $v_c$ was derived in
Eq.~\eqref{eq:critical velocity OLD}, leading to
\begin{equation}\label{eq:Mu_reducedOLD}
\tilde{\mu}=u\ell.
\end{equation}
We note that both in ND and OLD cases for large $N$ the work distribution
is described by Eqs.~\eqref{eq:Preduce}, \eqref{eq:Mu_reducedND} and
\eqref{eq:Mu_reducedOLD}, which for fixed dimensionless $u$ and $\ell$ are
independent of $N$. This is a direct consequence of the scaling of
internal relaxation times and internal length scales related to
scale-invariant internal structures of the polymer \cite{deGennes79}, when
only the polymer size $R$ and the largest relaxation time determine
the time and length scales of the internal modes.

Hydrodynamic interactions cannot be accurately treated even for ideal
polymers, since the equations of motion do not split into a set of independent
equations for each $q$ mode, as in Eq.~\eqref{eq:u_q_EOM}. However, it has
been shown \cite{Zimm56} that close to equilibrium such interactions can
be {\em mimicked} by replacing fixed $\gamma$ by a power law of $q$ in the
Fourier space, i.e., modifying $\tau_q$ in Eq.~\eqref{eq:tau_q} by an extra
power of $q$. [This change also requires a proper change in the noise correlation
$\langle \eta_{q}(t)\eta_{q'}(t')\rangle$.] While we expect these changes to
correctly reproduce the near-equilibrium behavior of the system, as well
as the value of $v_c$ in Eq.~\eqref{eq:v_c hydro}, we do not expect them
to produce a good approximation for $P(W)$ for $v\gg v_c$.

\section{Polymer near a wall}\label{sec: polymer near a wall}

\begin{figure}[t]
\includegraphics[width=9truecm]{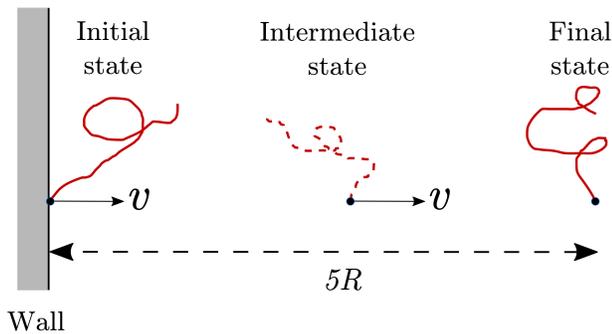}
\caption{(Color online) At $t=0$ a polymer is at equilibrium with one of
its ends attached to the wall. At $t>0$ it is dragged away from the wall
with a constant velocity $v$ until it reaches a distance $L=5R$, where
it is equilibrated. We consider a one-dimensional problem, and the
transverse dimension is only for illustration purposes.}
 \label{fig:Polymer_near wall}
\end{figure}

In this section we consider the Gaussian (Rouse) chain dragged away from a repulsive
wall.  At time $t=0$ the polymer is in equilibrium near the wall at $x_0=0$,
and is being dragged away from the wall at a constant velocity $v$, i.e.,
$x_0=vt$, as illustrated in Fig.~\ref{fig:Polymer_near wall}. If the final
distance $L$ of the polymer from the wall is significantly
larger than $R$, then in the final equilibrated state the free energy will
be equal to its value in free space, and in accordance with Eq.~\eqref{eq:Weq}
$\Delta F=-\frac{1}{2}k_BT\ln N$. Unlike the simple case
considered in the previous section, we can no longer expect a simple relation
between $\mu$ and $\sigma$, and $P(W)$ will not have a Gaussian form.

Our results rely on a numerical solution of Newton's equations in the ND case, and on a
solution of an overdamped Langevin equation in the OLD case \cite{Press92,Heermann90}.
In both cases
the calculation begins by choosing a properly weighted initial configuration.
The coordinates of the monomers are then advanced in time until $x_0$ reaches
the value $L=5R$. Integration of the external force that needs to be applied to $x_0$
to keep it moving at a constant velocity $v$ produces the work $W$ of the external
agent. Each calculation was repeated ${\cal N}=10^3$ times. Every time a new
initial equilibrium state was selected, and, in the OLD case a new thermal noise
function was generated. Such calculations produce a numerical estimate of $P(W)$
and can be used to produce a numerical estimate of $\Delta F$. We repeated the
calculations for chain sizes $N$ ranging from 10 to 100, and for each chain size
repeated the calculation for dragging velocities ranging from $0.1v_c$ to $5v_c$.
In this section we present a partial set of our results.

Since the exactly known $\Delta F$ is proportional to $-\ln N$, the graphs of
$P(W)$ shift towards more negative values of $W$ with increasing $N$.
For an easy comparison of the results with different chain sizes, it is convenient to
present all the functions in terms of the dissipated work $W_{\rm d}=W-\Delta F$, rather
than the entire work $W$. Furthermore, we will use the dimensionless variable
$y\equiv\beta W_d$. The probability density $\tilde{P}(y)$ is simply related to
$P(W)$ by $\tilde{P}(y)=P(W_{\rm d}+\Delta F)/\beta$. Similarly, a new shifted
function $\tilde{G}(y)\equiv e^{-y}\tilde{P}(y)$ can be used. In this notation
the probability density is normalized, i.e.,
$\int_{-\infty}^{\infty}\tilde{P}(y)\mathrm{d}y=1$ and the relation \eqref{eq:JE} has
the simple form $\int_{-\infty}^{\infty}\tilde{G}(y)\mathrm{d}y=1$. While these two
integrals impose some restrictions on the shape of $\tilde{P}$, there is still
plenty of room for dependence of this function on $N$ or $v$ and on the type
of dynamics (ND or OLD). Normally the term ``dissipation" implies positive
$W_d$ or $y>0$. In macroscopic systems for the average $W_{\rm d}$ this is referred
to as the Clausius inequality. However, a particular experiment may violate this
inequality \cite{Jarzynski11}. This is very unlikely, and it
can be shown that the probability of $y<-\zeta$ (for $\zeta>0$) is bounded by
$e^{-\zeta}$ \cite{Jarzynski11}. This means that $W_{\rm d}$ can be only a few times
$-k_BT$, independently of the system size. This inequality further restricts
the possible shapes of  $\tilde{P}$.

\begin{figure}[t]
\subfigure[ \label{sfig:NDa}]{%
  \includegraphics[width=9truecm]{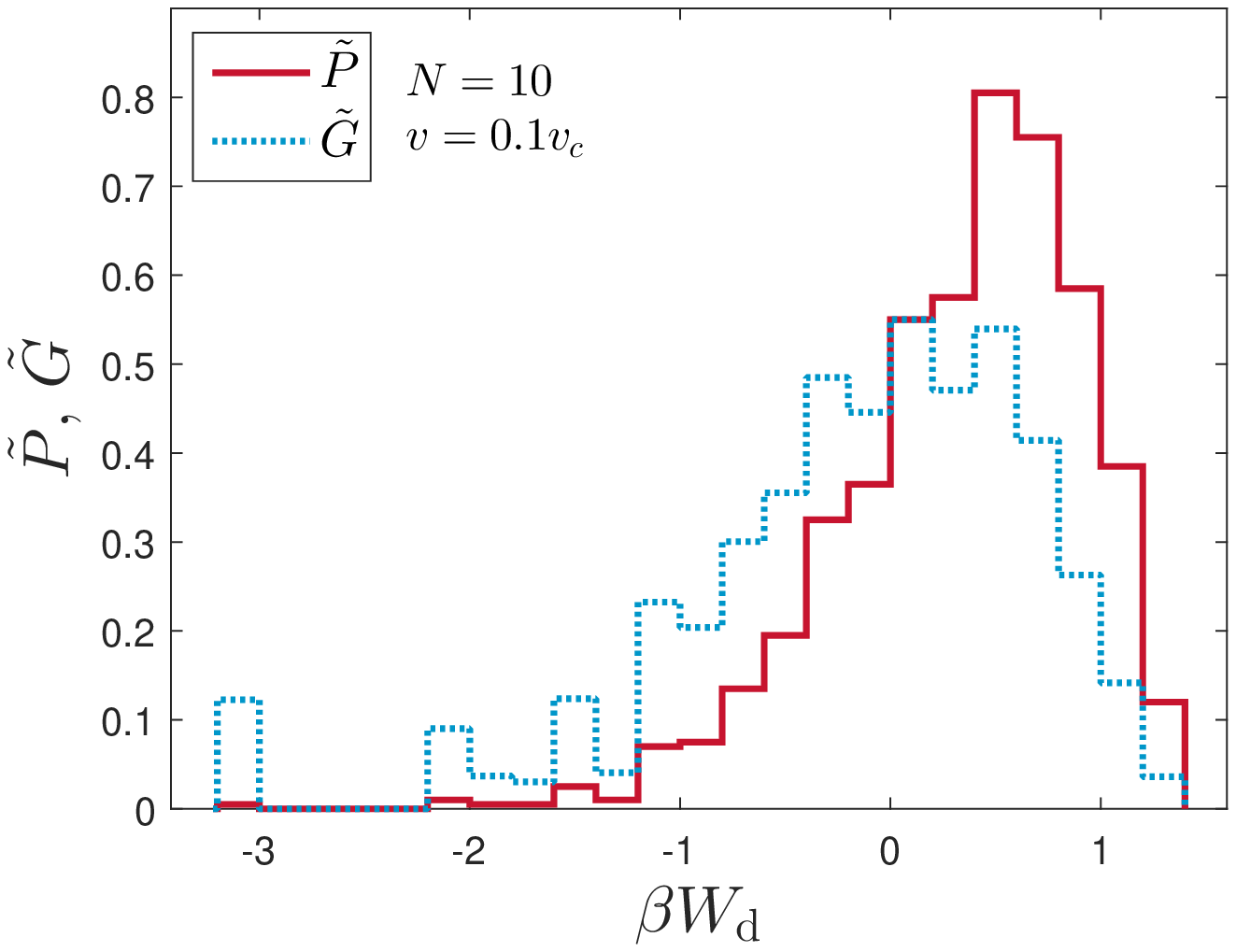}%
}\hfill
\subfigure[ \label{sfig:NDb}]{%
  \includegraphics[width=9truecm]{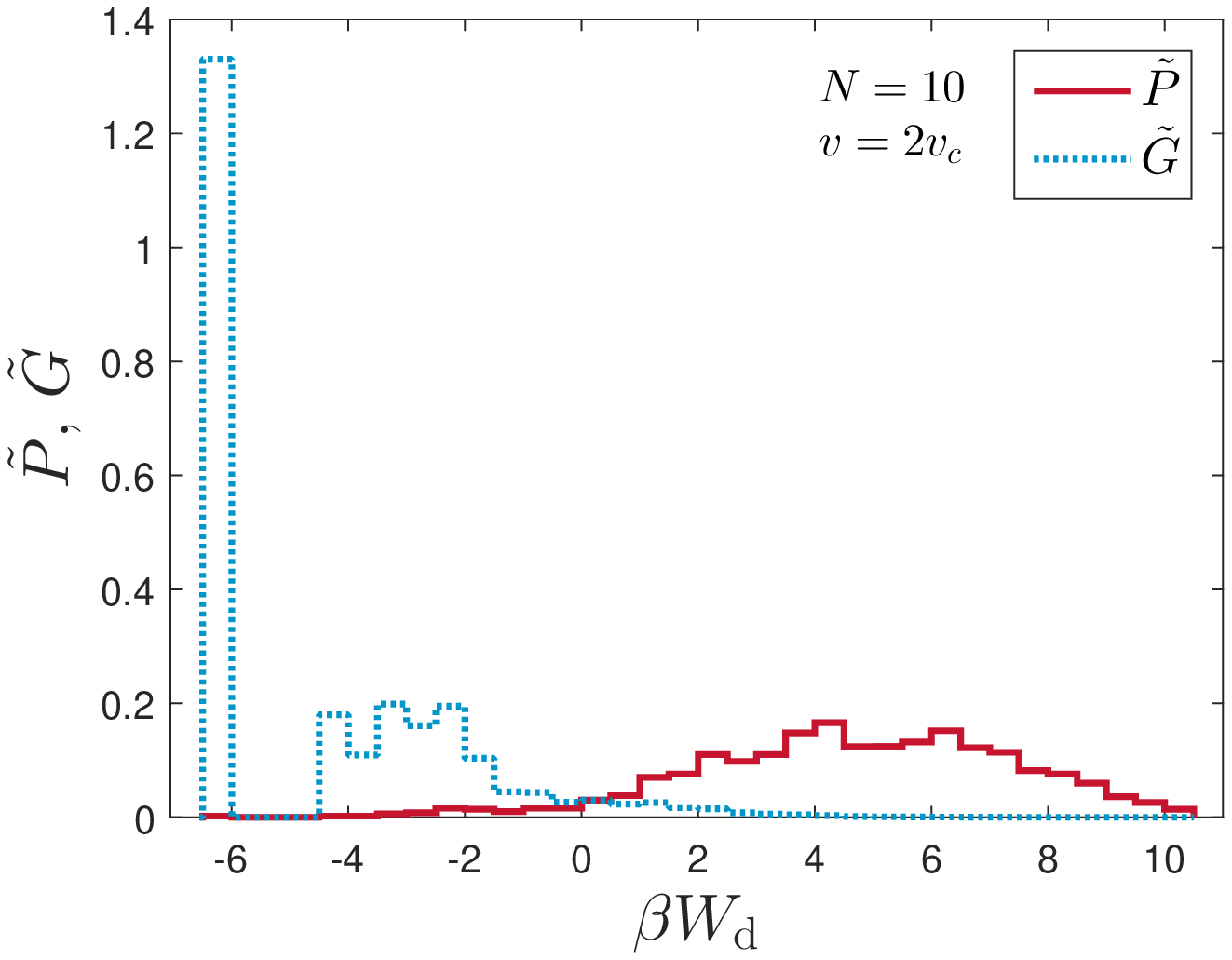}%
}
\caption{(Color online) Results for the dissipated work distribution in
the ND case extracted from a sample of ${\cal N}=10^3$ repeated
calculations for a short polymer ($N=10$) dragged away from a repulsive
wall. The solid line histograms depict the numerically calculated
probability density $\tilde{P}(y)$ measured as a function of
$y=\beta W_\mathrm{d}$, where  $W_\mathrm{d}$ is the dissipated work.
Dotted lines represent the shifted function $\tilde{G}(y)=e^{-y}\tilde{P}(y)$.
The polymer was dragged at a constant velocity (a) $v=0.1v_c$ and (b) $v=2v_c$.}
\label{fig: Pwd10_ND}
\end{figure}

\begin{figure}[t]
\subfigure[ \label{sfig:OLDa}]{%
  \includegraphics[width=0.5\textwidth]{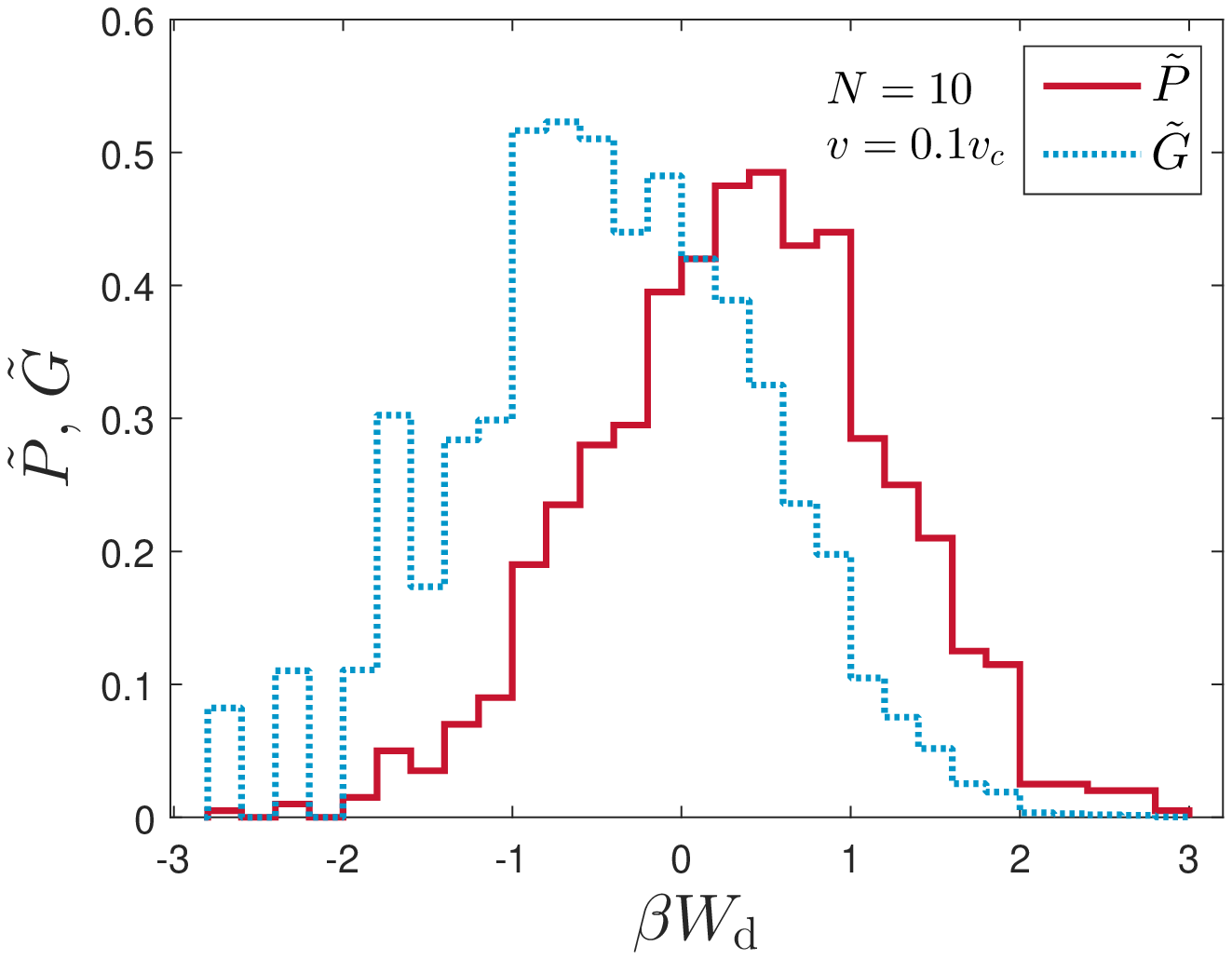}%
}\hfill
\subfigure[ \label{sfig:OLDb}]{%
  \includegraphics[width=0.5\textwidth]{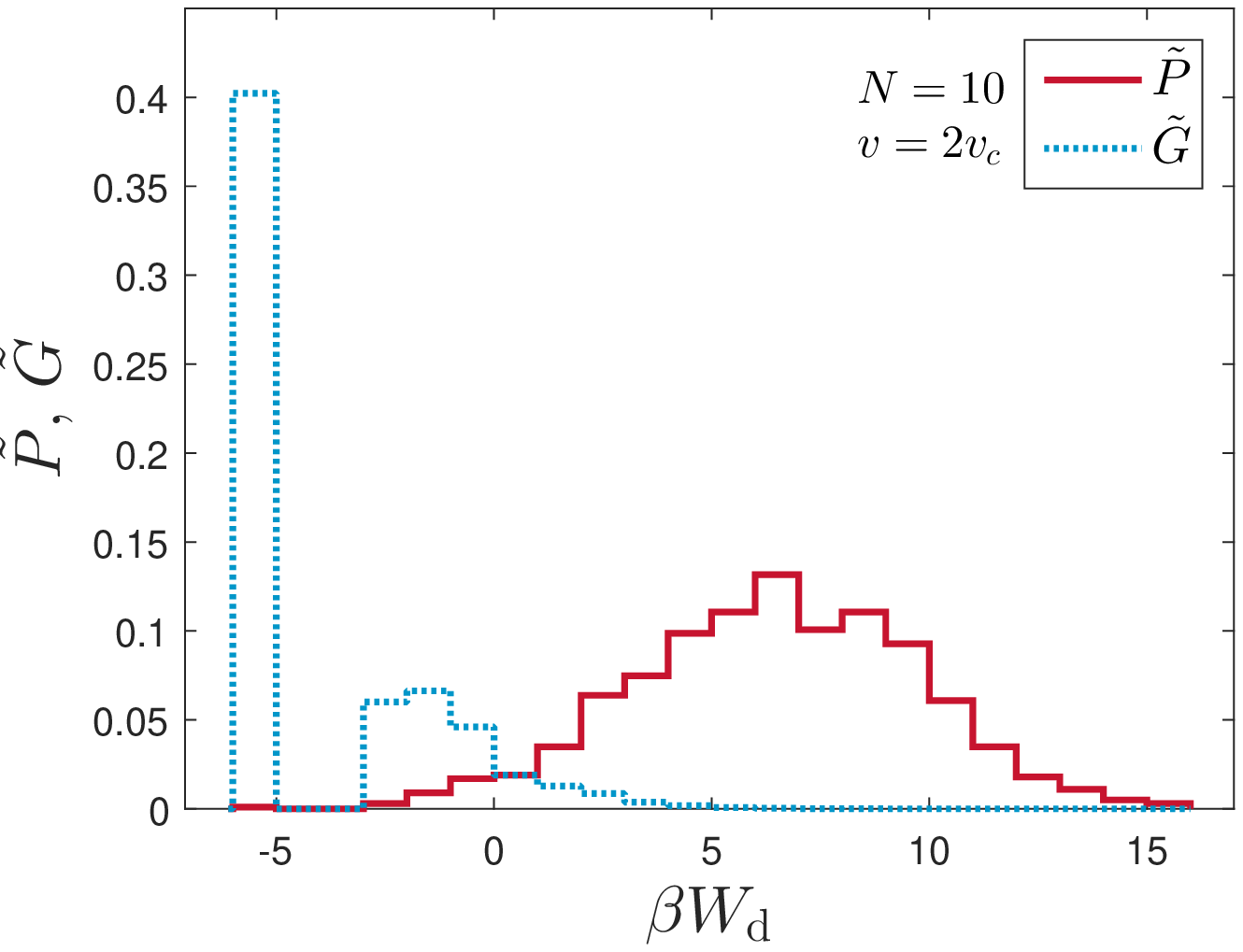}%
}
\caption{(Color online) Results for the dissipated work distribution in
the OLD case. All the notations are the same as in Fig.~\ref{fig: Pwd10_ND}.
}
\label{fig: Pwd10_OLD}
\end{figure}

The solid lines in Figs.~\ref{fig: Pwd10_ND} and \ref{fig: Pwd10_OLD} depict the
the probability distributions of the dissipated work $W_{\rm d}$ for short
polymers ($N=10$), for the ND and OLD cases, respectively. These histograms are
results of ${\cal N}=10^3$ repeated numerical experiments. The size of the
bin was selected for convenient presentation of the results. (Calculation of
$\Delta F$ from the data is performed directly from the set of measured
$W_{\rm d}$s rather than from these
histograms.) All distributions have a tail in the negative $y$ region but most
of such ``Clausius-inequality-violating" events are within one unit away from 0. For
small velocities $v=0.1v_c$ (graphs (a)) the distributions represent a process
that is rather close to quasistatic, i.e., they are narrow and close to 0, and
the mean dissipation satisfies
$\langle y\rangle=\beta \langle W_{\rm d}\rangle\ll 1$. When the
polymer is dragged away at a large velocity $v=2v_c$ [graphs (b)] the
distributions are shifted towards larger $y$ values, corresponding to an
external agent pulling the polymer away from the wall, in contrast with the
low-velocity case when the force is mostly towards the wall to maintain
a constant velocity.  The area  under the solid lines in all the graphs
is 1 due to normalization.

The dotted histograms in  Figs.~\ref{fig: Pwd10_ND} and \ref{fig: Pwd10_OLD}
represent the shifted functions $\tilde{G}$, and they were constructed
from the values of the solid histograms by multiplying them by $e^{-y}$.
To correctly reproduce the known value of $\Delta F$,  the area under
$\tilde{G}$ must be 1. This indeed happens in the low-velocity graphs (a),
where the area does not deviate from 1 by more that 0.1 even for a moderate
number $\cal N$ of experiments. At high velocities [graphs (b)] the
reconstructed $\tilde{G}$ is significantly shifted compared to $\tilde{P}$,
and is reconstructed from the poor quality tail of $\tilde{P}$. For $y<-1$
most bins correspond either to 0 or to 1 event found in that range, which
explains the noisy behavior of $\tilde{G}$. The area under the $\tilde{G}$
curve at high velocities significantly deviates from 1. This means that
the reconstructed free energy difference will have significant errors. We
used our data sets to directly evaluate the free energy difference from the
expression
\begin{equation}\label{eq: dFJ}
\Delta F_{\rm num}=-\frac{1}{\beta}\ln\left(\frac{1}{\cal N}
\sum_{i=1}^{\cal N}e^{-\beta W_{i}}\right),
\end{equation}
where $W_i$ is the work associated with the $i$th repetition of the calculation.
When these estimates of the free energy difference were compared with the
exactly known $\Delta F$, we saw (for $N$ ranging from 10 to 100) a fast
deterioration of accuracy when $v$ exceeded $v_c$. This result confirms our
expectation that $v_c$ serves as a borderline velocity between ``slow" and
``fast" processes for all values of $N$ in the problem of a polymer near a wall.

\begin{figure}[t!]
		\includegraphics[width=0.5\textwidth]{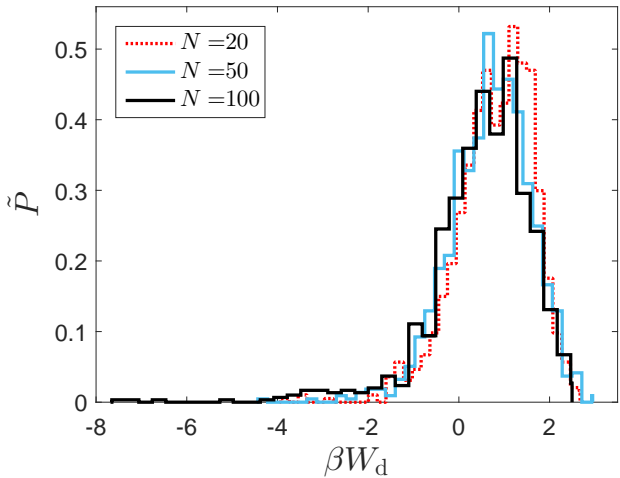}				
	\caption{(Color online) Results for the dissipated work distribution
in the ND case extracted from a sample of ${\cal N}=10^3$ repeated calculations
for polymers of three different $N$s as indicated in the legend.
Each calculation was performed with the same relative velocity $u=1$, or
$v=v_c$.}
\label{fig:PWscaled}
\end{figure}

In the previous section we found analytically that the work distribution
for a polymer of large $N$ in free space is described by
Eqs.~\eqref{eq:Preduce}, \eqref{eq:Mu_reducedND}, and \eqref{eq:Mu_reducedOLD},
which for fixed dimensionless $u$ and $\ell$ are independent of $N$.
We argued that in free space this is a direct consequence of the scaling of
internal relaxation times and internal length scales \cite{deGennes79}.
Such a lack of $N$ dependence is rather natural even in the presence of a wall,
since the equations of motion can be reformulated in properly scaled
variables in the $N\to\infty$ limit, indicating the existence of such a limit.
In free space, $\Delta F=0$, or $W_{\rm d}=W$. In the presence of the wall,
a simple free space result is no longer possible, due to the $N$ dependence of
$\Delta F$. Nevertheless, by considering $W_{\rm d}$ we eliminate the leading
$N$ dependence, and may hope to get an $N$-independent limit.
Figure \ref{fig:PWscaled} depicts $\tilde{P}(y)$ in the ND case for several values
of $N$, when each case has been calculated with the same {\em relative} velocity
$u$. Since $v_c$ decreases with increasing $N$, the velocity $v$ was also
decreased. We note that three different $N$s produce rather similar graphs.
A change of $u$ produces {\em different} graphs, but again, they seem to be
almost independent of $N$. In the third paragraph of this section we
mentioned that there were several constraints on the shape of $\tilde{P}(y)$.
Therefore, the similarity of the graphs is not surprising.
Nevertheless it is possible that in these scaled variables there is
an $N\to\infty$ limit of this graph which our numerical graphs are
approaching. We could see this property explicitly in the solution of
a polymer in free space. Due to scaling of the dynamical properties of
a polymer,  it is possible that similar features exist in the more
complicated case of a polymer near a wall. Similar behavior is also
observed in the OLD case, although the graphs for the same $u$ values
differ slightly from the ND shapes discussed above.

\section{Discussion}
\label{sec: Summary}

We studied the problem of a flexible polymer being dragged with
a constant velocity both in free space and in the vicinity of the wall, and
argued that there exists a critical $N$-dependent pulling velocity $v_c$
that separates the region of ``easy" reconstruction of $\Delta F$ by
means of the Jarzynski equality from the region of ``impossible"
reconstruction. The existence of a maximal deviation from equilibrium
for which the reconstruction of $\Delta F$ is possible is well known
from previous studies \cite{Liphardt02, Hummer05, Hummer05a, Harris07}.
In the context of unfolding of a large molecule this was typically viewed
as an event of a ``single particle" escaping from one well into
another \cite{Junier09},
or a sequence of such events \cite{Bustamante2000}.
We attempted to integrate the well-known static and dynamic scaling
properties of polymers into the description of their non-equilibrium
motion. Our heuristic argument in Sec.
\ref{sec:Determining v_c in polymers} produced in the ND case the result
$v_c\sim N^{-1/2}$ which was independent of the polymer type,
while for OLD the result depended on the exponent $\nu$.
The numerical support of our claim is limited to the two simple cases
of free space and repulsive wall for ideal polymers.  We observed the
lack of $N$-dependence of the dissipated work distribution, but the
range of $N$s was rather limited, and much longer polymers need to
be studied.

While our calculations were limited to the ideal polymers, some of
the concepts can be generalized to more realistic models, such as
the polymers in good solvents, when the interactions between non-adjacent
monomers are important. In the latter case, the Rouse modes, such as
$\tilde{x}_q$ in Eq.~\eqref{eq:DST x_n}, are no longer
the exact eigenmodes of the system. Nevertheless, the properly modified concept
of Rouse modes is used to describe dynamics of the polymers \cite{DoiM.Edwards1986}.
We expect that this and other generalized concepts can be used to produce results
described in  Sec.~\ref{sec:Determining v_c in polymers}. Confirmation of this
expectation will require detailed numerical simulations.

While the ``free draining" regime provides an adequate description of
the motion of polymers of moderate length, in experiments with
longer polymers the hydrodynamic interactions play an important role.
We briefly mentioned this regime in Sec.~\ref{sec:Determining v_c in polymers}.
In accordance with Eq.~\eqref{eq:v_c hydro} the typical critical
velocity of a 1$\mu$m size polymer will be slightly above
1 $\mu$m/s, which is one order of magnitude larger than the typical
speeds in many experiments (see, e.g., \cite{Liphardt02}).
The use of a constant dragging velocity
significantly simplified our derivations. By contrast, in real
experiments the force, rather than the speed, is controlled. However, in
homogeneous polymers these two should exhibit a similar behavior.

\begin{acknowledgments}
We thank Y.~Hammer and M.~Kardar for numerous discussions. We also thank
them and D.~J.~Bergman for valuable comments about the manuscript.
This work was supported by Israel Science Foundation Grant No.~453/17.
\end{acknowledgments}

\appendix
\section{Dragged harmonic oscillator} \label{APPEN:A}

Our analytical treatment of the Gaussian polymer in free space relies on
a decomposition into Rouse modes that are treated as simple dragged
harmonic oscillators. The harmonic oscillator (HO) was one of the first systems
used to demonstrate the workings of the JE \cite{Mazonka99}. The theoretical
treatment of a dragged HO \cite{Zon03a} followed an experimental study of
the translation of a particle in a harmonic optical trap \cite{Wang02}, which
was designed to test violations of the second law of thermodynamics with
findings consistent with a fluctuation theorem of Evans {\em et al.} \cite{Evans93}.

Consider the motion of a particle of mass $m$ at position $x$ attached
by a spring with force constant $k$ to a point $x_0$, which moves with
velocity $v$, i.e., at time $t$ its position is $x_0=vt$. In the ND case
the equation of motion is
\begin{equation}
m \ddot{x}=-k(x-vt),
\end{equation}
and its solution is given by
\begin{equation} \label{eq:x(t)_ND}
x(t)=x^0 \cos\left(\omega t\right)+\left(\frac{p^0}{m\omega}
-\frac{v}{\omega}\right)\sin\left(\omega t\right)+vt,
\end{equation}
where $\omega=\sqrt{k/m}$, and $x^0$, $p^0$ are the initial position
and momentum of the particle, respectively, which are selected from
a Gaussian distribution corresponding to the temperature of the system.
For the OLD case the equation of motion is
\begin{equation}
\gamma \dot{x}=-k(x-vt)+\eta (t),
\end{equation}
where $\gamma$ is the friction constant and random function $\eta(t)$
represents white Gaussian noise which satisfies $\langle\eta(t)\rangle=0$ and
$\left\langle \eta(t)\eta(t')\right\rangle
=2\gamma k_{{\rm B}}T\delta\left(t-t'\right)$.
The solution for this equation is given by
\begin{equation}
x(t)=x^0 e^{-t/\tau}+vt+\intop_{0}^{t}\left(\frac{1}{\gamma}\eta(t')
-v\right)e^{-(t-t')/\tau}\mathrm{d}t',
\end{equation}
where $\tau=\gamma/k$ is the relaxation time of the oscillator. Note,
that both in ND and in OLD cases the position $x(t)$ has a Gaussian
distribution since it is a linear combination of Gaussian variables.
The work done on the oscillator during the dragging of $x_0$ at a
constant velocity $v$ is
\begin{equation}
W =-v\intop_{0}^{t} k \left[x(t')-vt' \right] \rm{d} t'.
\end{equation}
Since $x(t)$ is known as function of the initial conditions and the realization
of noise $\eta(t)$ (in the OLD case), the distribution $P(W)$ can be easily
determined. Since $x(t)$ has a Gaussian distribution, the distribution of
work $P(W)$ is also a Gaussian characterized by its mean
$\mu = \left\langle W\right\rangle $ and variance
$\sigma^2 = \left\langle W^{2}\right\rangle
-\left\langle W\right\rangle ^{2}$.

Direct calculation of $\mu$ and $\sigma$, both in the ND and in the OLD cases, finds
that these quantities are simply related: $\mu(t)=\frac{\beta}{2}\sigma^2(t)$.
[This can also be viewed as a consequence of the JE for a Gaussian work
distribution in a situation where the equilibrium free energy of
an oscillator is independent of its position, as explained in the paragraph
following Eq.~\eqref{Eq:cumulant exp}.] In the ND case
\begin{equation} \label{eq:mean_var single ND}
\mu(t)= 2mv^{2}\sin^{2}\left(\frac{\omega t}{2}\right).
\end{equation}
This $\mu(t)$ is periodic and vanishes after each complete period of the oscillator.
In the OLD case
\begin{equation} \label{eq:mean_var single OLD}
\mu(t)=\gamma\tau v^{2}\left(e^{-t/\tau}+\frac{t}{\tau}-1\right).
\end{equation}
This $\mu(t)$ increases monotonically with time. At
large times the mean work is linear in $t$ representing the work against
friction.

\bibliographystyle{apsrev}

\begin{thebibliography}{39}
\expandafter\ifx\csname natexlab\endcsname\relax\def\natexlab#1{#1}\fi
\expandafter\ifx\csname bibnamefont\endcsname\relax
  \def\bibnamefont#1{#1}\fi
\expandafter\ifx\csname bibfnamefont\endcsname\relax
  \def\bibfnamefont#1{#1}\fi
\expandafter\ifx\csname citenamefont\endcsname\relax
  \def\citenamefont#1{#1}\fi
\expandafter\ifx\csname url\endcsname\relax
  \def\url#1{\texttt{#1}}\fi
\expandafter\ifx\csname urlprefix\endcsname\relax\def\urlprefix{URL }\fi
\providecommand{\bibinfo}[2]{#2}
\providecommand{\eprint}[2][]{\url{#2}}

\bibitem[{\citenamefont{Binder}(1983)}]{Binder83}
\bibinfo{author}{\bibfnamefont{K.}~\bibnamefont{Binder}}, in
  \emph{\bibinfo{booktitle}{Phase Transitions and Critical Phenomena}}, edited
  by \bibinfo{editor}{\bibfnamefont{C.}~\bibnamefont{Domb}} \bibnamefont{and}
  \bibinfo{editor}{\bibfnamefont{J.~L.} \bibnamefont{Lebowitz}}
  (\bibinfo{publisher}{Academic}, \bibinfo{address}{London},
  \bibinfo{year}{1983}), Vol.~\bibinfo{volume}{8}, pp. \bibinfo{pages}{1--144}.

\bibitem[{\citenamefont{De'Bell and Lookman}(1993)}]{DeBell93}
\bibinfo{author}{\bibfnamefont{K.}~\bibnamefont{De'Bell}} \bibnamefont{and}
  \bibinfo{author}{\bibfnamefont{T.}~\bibnamefont{Lookman}},
  \bibinfo{journal}{Rev. Mod. Phys.} \textbf{\bibinfo{volume}{65}},
  \bibinfo{pages}{87} (\bibinfo{year}{1993}).

\bibitem[{\citenamefont{Eisenriegler}(1993)}]{Eisenriegler1993}
\bibinfo{author}{\bibfnamefont{E.}~\bibnamefont{Eisenriegler}},
  \emph{\bibinfo{title}{Polymers Near Surfaces: Conformation Properties and
  Relation to Critical Phenomena}} (\bibinfo{publisher}{World Scientific},
  \bibinfo{address}{Singapore}, \bibinfo{year}{1993}).

\bibitem[{\citenamefont{de~Gennes}(1979)}]{deGennes79}
\bibinfo{author}{\bibfnamefont{P.~G.} \bibnamefont{de~Gennes}},
  \emph{\bibinfo{title}{Scaling Concepts in Polymer Physics}}
  (\bibinfo{publisher}{Cornell University Press}, \bibinfo{address}{Ithaca, NY}, 
  \bibinfo{year}{1979}).

\bibitem[{\citenamefont{Cardy and Redner}(1984)}]{Cardy84a}
\bibinfo{author}{\bibfnamefont{J.~L.} \bibnamefont{Cardy}} \bibnamefont{and}
  \bibinfo{author}{\bibfnamefont{S.}~\bibnamefont{Redner}},
  \bibinfo{journal}{J. Phys. A} \textbf{\bibinfo{volume}{17}},
  \bibinfo{pages}{L933} (\bibinfo{year}{1984}).

\bibitem[{\citenamefont{Grosberg et~al.}(1994)\citenamefont{Grosberg, Khokhlov,
  and Atanov}}]{Grosberg94}
\bibinfo{author}{\bibfnamefont{A.}~\bibnamefont{Grosberg}},
  \bibinfo{author}{\bibfnamefont{A.}~\bibnamefont{Khokhlov}}, \bibnamefont{and}
  \bibinfo{author}{\bibfnamefont{Y.}~\bibnamefont{Atanov}},
  \emph{\bibinfo{title}{Statistical Physics of Macromolecules}}
  (\bibinfo{publisher}{AIP}, \bibinfo{address}{New York},
  \bibinfo{year}{1994}).

\bibitem[{\citenamefont{Rubinstein and Colby}(2003)}]{Rubinstein03}
\bibinfo{author}{\bibfnamefont{M.}~\bibnamefont{Rubinstein}} \bibnamefont{and}
  \bibinfo{author}{\bibfnamefont{R.}~\bibnamefont{Colby}},
  \emph{\bibinfo{title}{Polymer Physics}} (\bibinfo{publisher}{Oxford
  University Press}, \bibinfo{year}{2003}).

\bibitem[{\citenamefont{Zlatanova and van Holde}(2006)}]{Zlatanova06}
\bibinfo{author}{\bibfnamefont{J.}~\bibnamefont{Zlatanova}} \bibnamefont{and}
  \bibinfo{author}{\bibfnamefont{K.}~\bibnamefont{van Holde}},
  \bibinfo{journal}{Mol. Cell} \textbf{\bibinfo{volume}{24}},
  \bibinfo{pages}{317 } (\bibinfo{year}{2006}).

\bibitem[{\citenamefont{Leuba and Zlatanova}(2001)}]{Leuba01}
\bibinfo{author}{\bibfnamefont{S.}~\bibnamefont{Leuba}} \bibnamefont{and}
  \bibinfo{author}{\bibfnamefont{J.}~\bibnamefont{Zlatanova}},
  \emph{\bibinfo{title}{Biology at the Single Molecule Level}}
  (\bibinfo{publisher}{Pergamon}, \bibinfo{address}{New York},
  \bibinfo{year}{2001}).

\bibitem[{\citenamefont{Binnig et~al.}(1986)\citenamefont{Binnig, Quate, and
  Gerber}}]{Binnig86}
\bibinfo{author}{\bibfnamefont{G.}~\bibnamefont{Binnig}},
  \bibinfo{author}{\bibfnamefont{C.~F.} \bibnamefont{Quate}}, \bibnamefont{and}
  \bibinfo{author}{\bibfnamefont{C.}~\bibnamefont{Gerber}},
  \bibinfo{journal}{Phys. Rev. Lett.} \textbf{\bibinfo{volume}{56}},
  \bibinfo{pages}{930} (\bibinfo{year}{1986}).

\bibitem[{\citenamefont{Morita et~al.}(2002)\citenamefont{Morita, Wiesendanger,
  and Meyer}}]{Morita02}
\bibinfo{author}{\bibfnamefont{S.}~\bibnamefont{Morita}},
  \bibinfo{author}{\bibfnamefont{R.}~\bibnamefont{Wiesendanger}},
  \bibnamefont{and} \bibinfo{author}{\bibfnamefont{E.}~\bibnamefont{Meyer}},
  \emph{\bibinfo{title}{Noncontact Atomic Force Microscopy}},
  vol.~\bibinfo{volume}{1} (\bibinfo{publisher}{Springer},
  \bibinfo{address}{New York}, \bibinfo{year}{2002}).

\bibitem[{\citenamefont{Sarid}(1994)}]{Sarid94}
\bibinfo{author}{\bibfnamefont{D.}~\bibnamefont{Sarid}},
  \emph{\bibinfo{title}{Scanning Force Microscopy}} (\bibinfo{publisher}{Oxford
  University Press}, \bibinfo{year}{1994}).

\bibitem[{\citenamefont{Bustamante et~al.}(2000)\citenamefont{Bustamante,
  Smith, Liphardt, and Smith}}]{Bustamante2000}
\bibinfo{author}{\bibfnamefont{C.}~\bibnamefont{Bustamante}},
  \bibinfo{author}{\bibfnamefont{S.~B.} \bibnamefont{Smith}},
  \bibinfo{author}{\bibfnamefont{J.}~\bibnamefont{Liphardt}}, \bibnamefont{and}
  \bibinfo{author}{\bibfnamefont{D.}~\bibnamefont{Smith}},
  \bibinfo{journal}{Curr. Opin. Struc. Biol.} \textbf{\bibinfo{volume}{10}},
  \bibinfo{pages}{279} (\bibinfo{year}{2000}).

\bibitem[{\citenamefont{Maghrebi et~al.}(2011)\citenamefont{Maghrebi, Kantor,
  and Kardar}}]{Maghrebi11}
\bibinfo{author}{\bibfnamefont{M.~F.} \bibnamefont{Maghrebi}},
  \bibinfo{author}{\bibfnamefont{Y.}~\bibnamefont{Kantor}}, \bibnamefont{and}
  \bibinfo{author}{\bibfnamefont{M.}~\bibnamefont{Kardar}},
  \bibinfo{journal}{Europhys. Lett.} \textbf{\bibinfo{volume}{96}},
  \bibinfo{pages}{66002} (\bibinfo{year}{2011}).

\bibitem[{\citenamefont{Maghrebi et~al.}(2012)\citenamefont{Maghrebi, Kantor,
  and Kardar}}]{Maghrebi12}
\bibinfo{author}{\bibfnamefont{M.~F.} \bibnamefont{Maghrebi}},
  \bibinfo{author}{\bibfnamefont{Y.}~\bibnamefont{Kantor}}, \bibnamefont{and}
  \bibinfo{author}{\bibfnamefont{M.}~\bibnamefont{Kardar}},
  \bibinfo{journal}{Phys. Rev. E} \textbf{\bibinfo{volume}{86}},
  \bibinfo{pages}{061801} (\bibinfo{year}{2012}).

\bibitem[{\citenamefont{Alfasi and Kantor}(2015)}]{AK_PRE91}
\bibinfo{author}{\bibfnamefont{N.}~\bibnamefont{Alfasi}} \bibnamefont{and}
  \bibinfo{author}{\bibfnamefont{Y.}~\bibnamefont{Kantor}},
  \bibinfo{journal}{Phys. Rev. E} \textbf{\bibinfo{volume}{91}},
  \bibinfo{pages}{042126} (\bibinfo{year}{2015}).

\bibitem[{\citenamefont{Hammer and Kantor}(2015)}]{HK_PRE92}
\bibinfo{author}{\bibfnamefont{Y.}~\bibnamefont{Hammer}} \bibnamefont{and}
  \bibinfo{author}{\bibfnamefont{Y.}~\bibnamefont{Kantor}},
  \bibinfo{journal}{Phys. Rev. E} \textbf{\bibinfo{volume}{92}},
  \bibinfo{pages}{062602} (\bibinfo{year}{2015}).

\bibitem[{\citenamefont{Jarzynski}(1997{\natexlab{a}})}]{Jarzynski97a}
\bibinfo{author}{\bibfnamefont{C.}~\bibnamefont{Jarzynski}},
  \bibinfo{journal}{Phys. Rev. Lett.} \textbf{\bibinfo{volume}{78}},
  \bibinfo{pages}{2690} (\bibinfo{year}{1997}{\natexlab{a}}).

\bibitem[{\citenamefont{Jarzynski}(1997{\natexlab{b}})}]{Jarzynski97b}
\bibinfo{author}{\bibfnamefont{C.}~\bibnamefont{Jarzynski}},
  \bibinfo{journal}{Phys. Rev. E} \textbf{\bibinfo{volume}{56}},
  \bibinfo{pages}{05018} (\bibinfo{year}{1997}{\natexlab{b}}).

\bibitem[{\citenamefont{Liphardt et~al.}(2002)\citenamefont{Liphardt, Dumont,
  Smith, {Tinoco Jr.}, and Bustamante}}]{Liphardt02}
\bibinfo{author}{\bibfnamefont{J.}~\bibnamefont{Liphardt}},
  \bibinfo{author}{\bibfnamefont{S.}~\bibnamefont{Dumont}},
  \bibinfo{author}{\bibfnamefont{S.~B.} \bibnamefont{Smith}},
  \bibinfo{author}{\bibfnamefont{I.}~\bibnamefont{{Tinoco Jr.}}},
  \bibnamefont{and}
  \bibinfo{author}{\bibfnamefont{C.}~\bibnamefont{Bustamante}},
  \bibinfo{journal}{Science} \textbf{\bibinfo{volume}{296}},
  \bibinfo{pages}{1832} (\bibinfo{year}{2002}).

\bibitem[{\citenamefont{Hummer and Szabo}(2005{\natexlab{a}})}]{Hummer05}
\bibinfo{author}{\bibfnamefont{G.}~\bibnamefont{Hummer}} \bibnamefont{and}
  \bibinfo{author}{\bibfnamefont{A.}~\bibnamefont{Szabo}},
  \bibinfo{journal}{Acc. Chem. Res.} \textbf{\bibinfo{volume}{38}},
  \bibinfo{pages}{504} (\bibinfo{year}{2005}{\natexlab{a}}).

\bibitem[{\citenamefont{Hummer and Szabo}(2005{\natexlab{b}})}]{Hummer05a}
\bibinfo{author}{\bibfnamefont{G.}~\bibnamefont{Hummer}} \bibnamefont{and}
  \bibinfo{author}{\bibfnamefont{A.}~\bibnamefont{Szabo}},
  \bibinfo{journal}{Proc. Natl. Acad. Sci. USA} \textbf{\bibinfo{volume}{98}}, \bibinfo{pages}{3658}
  (\bibinfo{year}{2001}{\natexlab{b}}).

\bibitem[{\citenamefont{Harris et~al.}(2007)\citenamefont{Harris, Song, and
  Kiang}}]{Harris07}
\bibinfo{author}{\bibfnamefont{N.~C.} \bibnamefont{Harris}},
  \bibinfo{author}{\bibfnamefont{Y.}~\bibnamefont{Song}}, \bibnamefont{and}
  \bibinfo{author}{\bibfnamefont{C.-H.} \bibnamefont{Kiang}},
  \bibinfo{journal}{Phys. Rev. Lett.} \textbf{\bibinfo{volume}{99}},
  \bibinfo{pages}{068101} (\bibinfo{year}{2007}).

\bibitem[{\citenamefont{Jarzynski}(2011)}]{Jarzynski11}
\bibinfo{author}{\bibfnamefont{C.}~\bibnamefont{Jarzynski}},
  \bibinfo{journal}{Annu. Rev. Condens. Matter Phys.}
  \textbf{\bibinfo{volume}{2}}, \bibinfo{pages}{329} (\bibinfo{year}{2011}).

\bibitem[{\citenamefont{Lua and Grosberg}(2004)}]{Lua04}
\bibinfo{author}{\bibfnamefont{R.~C.} \bibnamefont{Lua}} \bibnamefont{and}
  \bibinfo{author}{\bibfnamefont{A.~Y.} \bibnamefont{Grosberg}},
  \bibinfo{journal}{J. Chem. Phys. B} \textbf{\bibinfo{volume}{109}},
  \bibinfo{pages}{6805} (\bibinfo{year}{2005}).

\bibitem[{\citenamefont{Bena et~al.}(2005)\citenamefont{Bena, van~den Broeck,
  and Kawai}}]{Bena05}
\bibinfo{author}{\bibfnamefont{I.}~\bibnamefont{Bena}},
  \bibinfo{author}{\bibfnamefont{C.}~\bibnamefont{van~den Broeck}},
  \bibnamefont{and} \bibinfo{author}{\bibfnamefont{R.}~\bibnamefont{Kawai}},
  \bibinfo{journal}{Europhys. Lett.} \textbf{\bibinfo{volume}{71}},
  \bibinfo{pages}{879} (\bibinfo{year}{2005}).

\bibitem[{\citenamefont{Jarzynski}(2006)}]{Jarzynski06}
\bibinfo{author}{\bibfnamefont{C.}~\bibnamefont{Jarzynski}},
  \bibinfo{journal}{Phys. Rev. E} \textbf{\bibinfo{volume}{73}},
  \bibinfo{pages}{046105} (\bibinfo{year}{2006}).

\bibitem[{\citenamefont{Doi and Edwards}(1986)}]{DoiM.Edwards1986}
\bibinfo{author}{\bibfnamefont{M.}~\bibnamefont{Doi}} \bibnamefont{and}
  \bibinfo{author}{\bibfnamefont{S.}~\bibnamefont{Edwards}},
  \emph{\bibinfo{title}{{The Theory of Polymer Dynamics}}}
  (\bibinfo{publisher}{Oxford University Press, New York}, \bibinfo{year}{1986}).

\bibitem[{\citenamefont{Zimm}(1956)}]{Zimm56}
\bibinfo{author}{\bibfnamefont{B.~H.} \bibnamefont{Zimm}}, \bibinfo{journal}{J.
  Chem. Phys.} \textbf{\bibinfo{volume}{24}}, \bibinfo{pages}{269}
  (\bibinfo{year}{1956}).

\bibitem[{\citenamefont{Dhar}(2005)}]{Dhar05}
\bibinfo{author}{\bibfnamefont{A.}~\bibnamefont{Dhar}}, \bibinfo{journal}{Phys.
  Rev. E} \textbf{\bibinfo{volume}{71}}, \bibinfo{pages}{036126}
  (\bibinfo{year}{2005}).

\bibitem[{\citenamefont{Speck and Seifert}(2005)}]{Speck05}
\bibinfo{author}{\bibfnamefont{T.}~\bibnamefont{Speck}} \bibnamefont{and}
  \bibinfo{author}{\bibfnamefont{U.}~\bibnamefont{Seifert}},
  \bibinfo{journal}{Eur. J. Phys. B} \textbf{\bibinfo{volume}{43}},
  \bibinfo{pages}{521} (\bibinfo{year}{2005}).

\bibitem[{\citenamefont{Hummer}(2001)}]{Hummer2001}
\bibinfo{author}{\bibfnamefont{G.}~\bibnamefont{Hummer}},
  \bibinfo{journal}{J. Chem. Phys.} \textbf{\bibinfo{volume}{114}},
  \bibinfo{pages}{7330} (\bibinfo{year}{2001}).

\bibitem[{\citenamefont{Press et~al.}(1992)\citenamefont{Press, Teukolsky,
  Vetterling, and Flannery}}]{Press92}
\bibinfo{author}{\bibfnamefont{W.~H.} \bibnamefont{Press}},
  \bibinfo{author}{\bibfnamefont{S.~A.} \bibnamefont{Teukolsky}},
  \bibinfo{author}{\bibfnamefont{W.~T.} \bibnamefont{Vetterling}},
  \bibnamefont{and} \bibinfo{author}{\bibfnamefont{B.~P.}
  \bibnamefont{Flannery}}, \emph{\bibinfo{title}{Numerical recipes in C}}
  (\bibinfo{publisher}{Cambridge University Press, Cambridge, England}, \bibinfo{year}{1992}).

\bibitem[{\citenamefont{Heermann}(1990)}]{Heermann90}
\bibinfo{author}{\bibfnamefont{D.~W.} \bibnamefont{Heermann}},
  \emph{\bibinfo{title}{Computer Simulation Methods in Theoretical Physics}}
  (\bibinfo{publisher}{Springer}, \bibinfo{address}{Berlin},
  \bibinfo{year}{1990}).

\bibitem[{\citenamefont{Junier et~al.}(2009)\citenamefont{Junier, Mossa,
  Manosas, and Ritort}}]{Junier09}
\bibinfo{author}{\bibfnamefont{I.}~\bibnamefont{Junier}},
  \bibinfo{author}{\bibfnamefont{A.}~\bibnamefont{Mossa}},
  \bibinfo{author}{\bibfnamefont{M.}~\bibnamefont{Manosas}}, \bibnamefont{and}
  \bibinfo{author}{\bibfnamefont{F.}~\bibnamefont{Ritort}},
  \bibinfo{journal}{Phys. Rev. Lett.} \textbf{\bibinfo{volume}{102}},
  \bibinfo{pages}{070602} (\bibinfo{year}{2009}).

\bibitem[{\citenamefont{Mazonka and Jarzynski}(1999)}]{Mazonka99}
\bibinfo{author}{\bibfnamefont{O.}~\bibnamefont{Mazonka}} \bibnamefont{and}
  \bibinfo{author}{\bibfnamefont{C.}~\bibnamefont{Jarzynski}},
  \bibinfo{journal}{arXiv:condmat/991212}.

\bibitem[{\citenamefont{van Zon and Cohen}(2003)}]{Zon03a}
\bibinfo{author}{\bibfnamefont{R.}~\bibnamefont{van Zon}} \bibnamefont{and}
  \bibinfo{author}{\bibfnamefont{E.~G.~D.} \bibnamefont{Cohen}},
  \bibinfo{journal}{Phys. Rev. E} \textbf{\bibinfo{volume}{67}},
  \bibinfo{pages}{046102} (\bibinfo{year}{2003}).

\bibitem[{\citenamefont{Wang et~al.}(2002)\citenamefont{Wang, Sevick, Mittag,
  Searles, and Evans}}]{Wang02}
\bibinfo{author}{\bibfnamefont{G.~M.} \bibnamefont{Wang}},
  \bibinfo{author}{\bibfnamefont{E.~M.} \bibnamefont{Sevick}},
  \bibinfo{author}{\bibfnamefont{E.}~\bibnamefont{Mittag}},
  \bibinfo{author}{\bibfnamefont{D.~J.} \bibnamefont{Searles}},
  \bibnamefont{and} \bibinfo{author}{\bibfnamefont{D.~J.} \bibnamefont{Evans}},
  \bibinfo{journal}{Phys. Rev. Lett.} \textbf{\bibinfo{volume}{89}},
  \bibinfo{pages}{50601} (\bibinfo{year}{2002}).

\bibitem[{\citenamefont{Evans et~al.}(1993)\citenamefont{Evans, Cohen, and
  Morriss}}]{Evans93}
\bibinfo{author}{\bibfnamefont{D.~J.} \bibnamefont{Evans}},
  \bibinfo{author}{\bibfnamefont{E.~G.~D.} \bibnamefont{Cohen}},
  \bibnamefont{and} \bibinfo{author}{\bibfnamefont{G.~P.}
  \bibnamefont{Morriss}}, \bibinfo{journal}{Phys. Rev. Lett.}
  \textbf{\bibinfo{volume}{71}}, \bibinfo{pages}{2401} (\bibinfo{year}{1993}).

\end{thebibliography}

\end{document}